\documentclass[a4paper,11pt]{article}
\usepackage{pos}

\title{Baryon and Meson Excited States}
\ShortTitle{Baryon and Meson Excited States...}

\author[a]{L. David Roper}
\author*[b]{Igor Strakovsky}

\affiliation[a]{Prof. Emeritus of Physics, Virginia Polytechnic Institute and State University\\ 1001 Auburn Dr. SW, Blacksburg VA 24060, USA}

\affiliation[b]{Institute for Nuclear Studies, Department of Physics,\\
  The George Washington University, Washington, DC 20052, USA}

\emailAdd{roperld@vt.edu}
\emailAdd{igor@gwu.edu}

\abstract{The masses of fifteen baryon sets and twenty-four meson sets of three or more equal-quantum excited states are fitted by a simple two-parameter logarithm function, $M_n = \alpha Ln(n) + \beta$, where $n$ is the level of radial excitation. The conjecture is made that accurately measured masses using Breat-Wigner PDG2024 data at fixed $J^P$ for baryons and $J^{PC}$ for mesons~\cite{ParticleDataGroup:2024cfk} of all equal-quantum baryons (including LHCb exotic $P_{c\bar{c}}^+$s) and meson (including $s\bar{s}$, $s\bar{c}$, $c\bar{c}$, $c\bar{b}$, and $b\bar{b}$) excited states are related by the logarithm function used here; at least for the mass range of currently known excited states. Thus, a universal mass equation for equal-quantum excited-states sets is 
presented~\cite{Roper:2024ovj}. The masses of twelve baryon sets and sixteen meson sets, with only two equal-quantum excited states in each set, using Breit-Wigner PDG2024 masses and their uncertainties, are fitted by thr universal mass equation~\cite{Roper:2025ldj}.
}

\FullConference{%
  The 21st International Conference on Hadron Spectroscopy and Structure - HADRON2025 \\
  27-31 March, 2025 \\ 
  Toyonaka Campus, Osaka University, Osaka, Japan
}

\begin{document}
\maketitle

\section{Introduction}
This research started as a study of the excited states of the proton/neutron. The interesting result of that study led to studies of other equal-quantum combinations of particle-physics resonances. 

Data for baryon at fixed $J^P$  and meson at fixed $J^{PC}$ excited states are reported in the Particle Data Listings from the Particle Data Group 
(PDG)~\cite{ParticleDataGroup:2024cfk}. Resonance mass and width, with their uncertainties, are reported as Breit-Wigner (BW) values.

A universal mass equation (UME) is presented for equal-quantum excited-state sets.

\section{Universal Mass Equation}
In plotting the mass data for the $N1/2^+$ data set, we found an excellent fit at fixed $J^P$ for baryons and fixed $J^{PC}$ for mesons, using a $\chi^2$ minimization procedure for the logarithm function or UME for Equal-Quantum Excited-States sets
\begin{equation}
    M_n = \alpha~Ln(n) + \beta \>,
\label{eq:eq1}
\end{equation}
and used the fit to predict the masses of one missing excited state and four higher excited states of the proton/neutron~\cite{Roper:2024ovj}. Here, $n$ is the radial excitation level and $\alpha$ (logarithmic slope) and $\beta$ (essentially ground mass in data set) are free parameters. (The parameter $\beta = M_1$, since $Ln(1) = 0$, especially if the lowest mass is the mass measured most accurately of the set, as the neutron is in the data set $N1/2^+$.) 

An excited state is a quantum state of a system that has a higher energy than the ground state $M_1$. In this work, the authors often label a ground state as an excited state of the vacuum. Our Part~II~\cite{Roper:2025ldj} reports using data sets with only two known excited states with masses M\textsubscript{1} and M\textsubscript{2} (duo sets), which have reasonably low measurement uncertainties. Here the parameter $\beta = M_1$.

%
%
%
%
%

\section{UME Samples for Baryons}
\begin{figure}[htb!]
    \centering
    \includegraphics[width=0.45\linewidth]{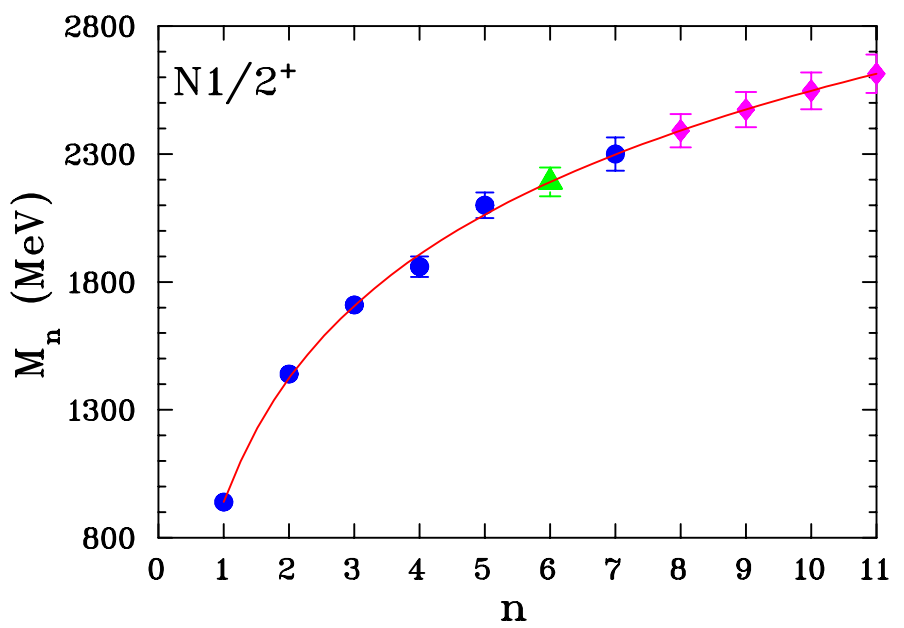}~~~
    \includegraphics[width=0.45\linewidth]{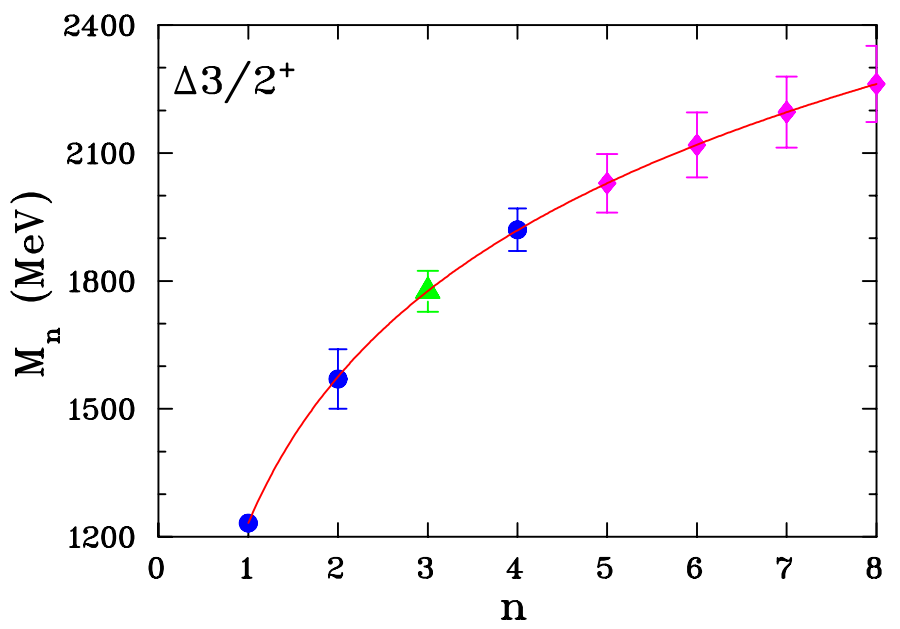}
    \caption{$N1/2^+$ set (left) and $\Delta 3/2^+$ set (right).
    PDG2024 data (blue circles)~\cite{ParticleDataGroup:2024cfk}.
    The green triangle is the calculated mass of the missing states.
    Predicted states (magenta diamonds).
    The solid red curve presents the best-fit. Part~I~\cite{Roper:2024ovj}.
    }
    \label{fig:N1/2+}
\end{figure}

\begin{figure}[htb!]
    \centering
    \includegraphics[width=0.4\linewidth]{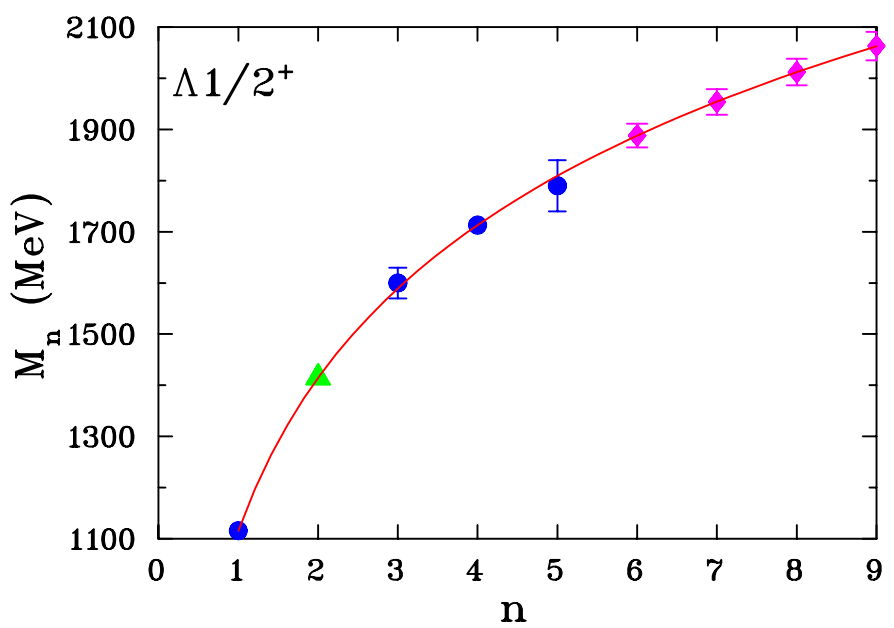}~~~
    \includegraphics[width=0.4\linewidth]{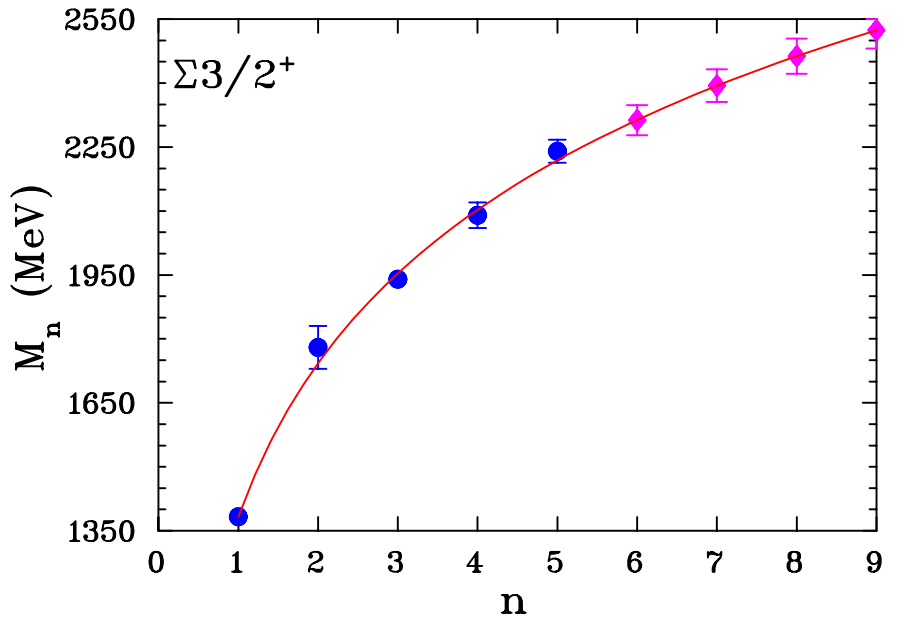}
    \caption{$\Lambda 1/2^+$ set (left) and $\Sigma 3/2^+$ set (right). 
    The notation is the same as in Fig.~\ref{fig:N1/2+}.
    Part~I~\cite{Roper:2024ovj}
    }
    \label{fig:Lambda1/2+}
\end{figure}

\begin{figure}[htb!]
    \centering
    \includegraphics[width=0.4\linewidth]{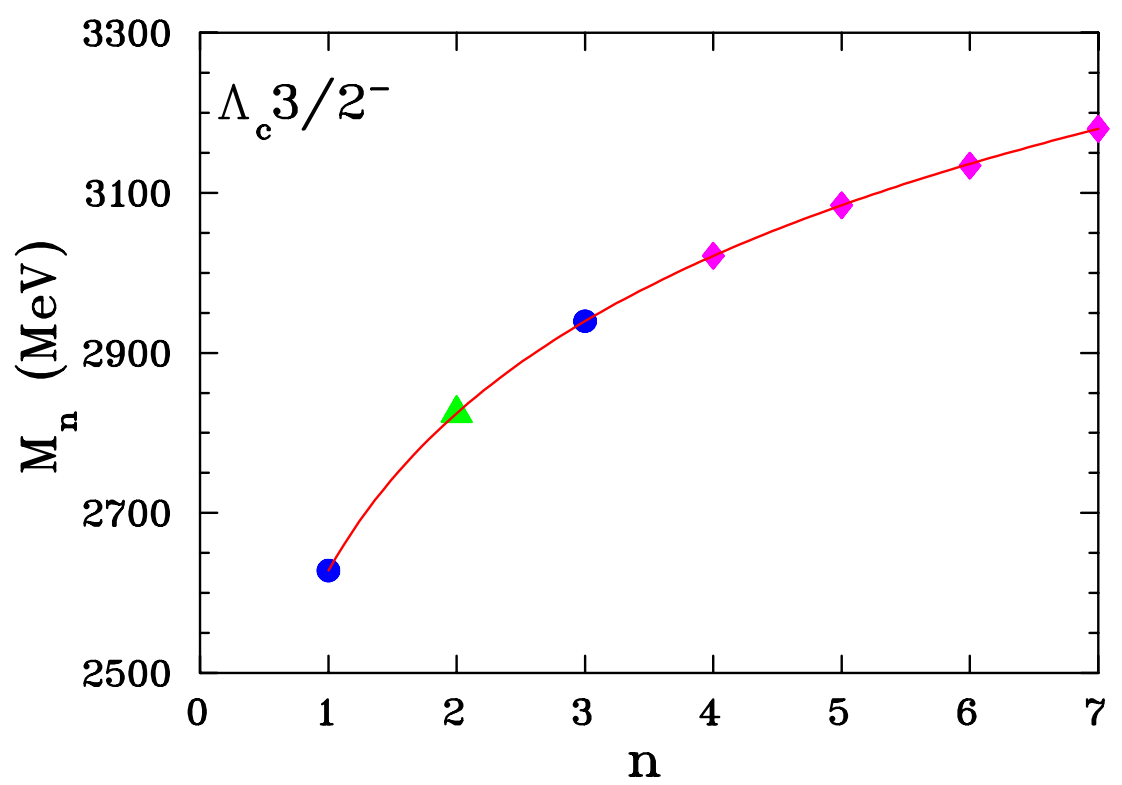}~~~
    \includegraphics[width=0.4\linewidth]{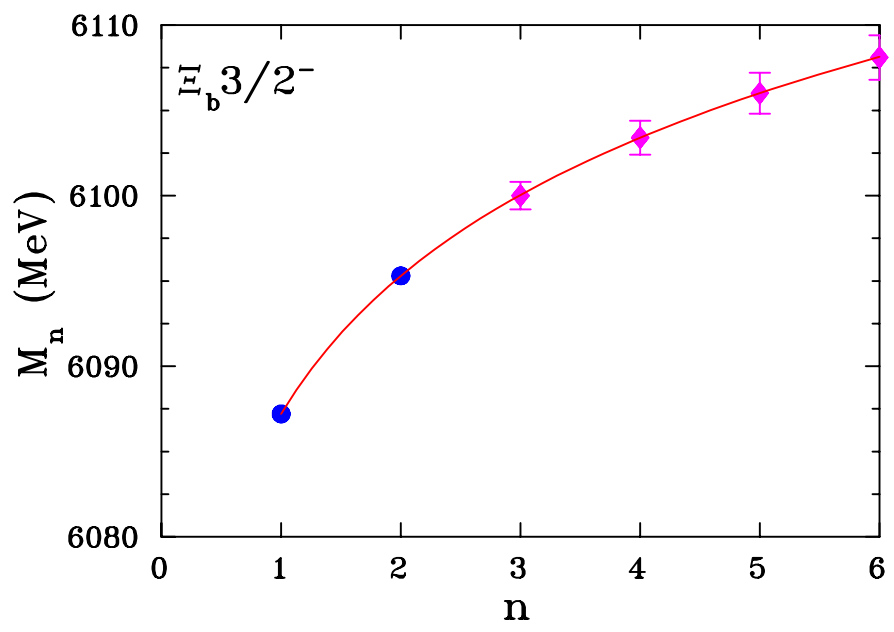}
    \caption{$\Lambda_c3/2^-$ duo set (left) and $\Xi_b3/2^-$ duo set (right). 
    The notation is the same as in Fig.~\ref{fig:N1/2+}.
    Part~II~\cite{Roper:2025ldj}.
    }
    \label{fig:Lambdac3/2-}
\end{figure}

Samples for the Baryon case are given in Figs.~\ref{fig:N1/2+} through \ref{fig:Lambdac3/2-}.

\subsection{Systematics of Baryon Excited-States Sets}
\begin{figure}[htb!]
    \centering
    \includegraphics[width=0.5\linewidth]{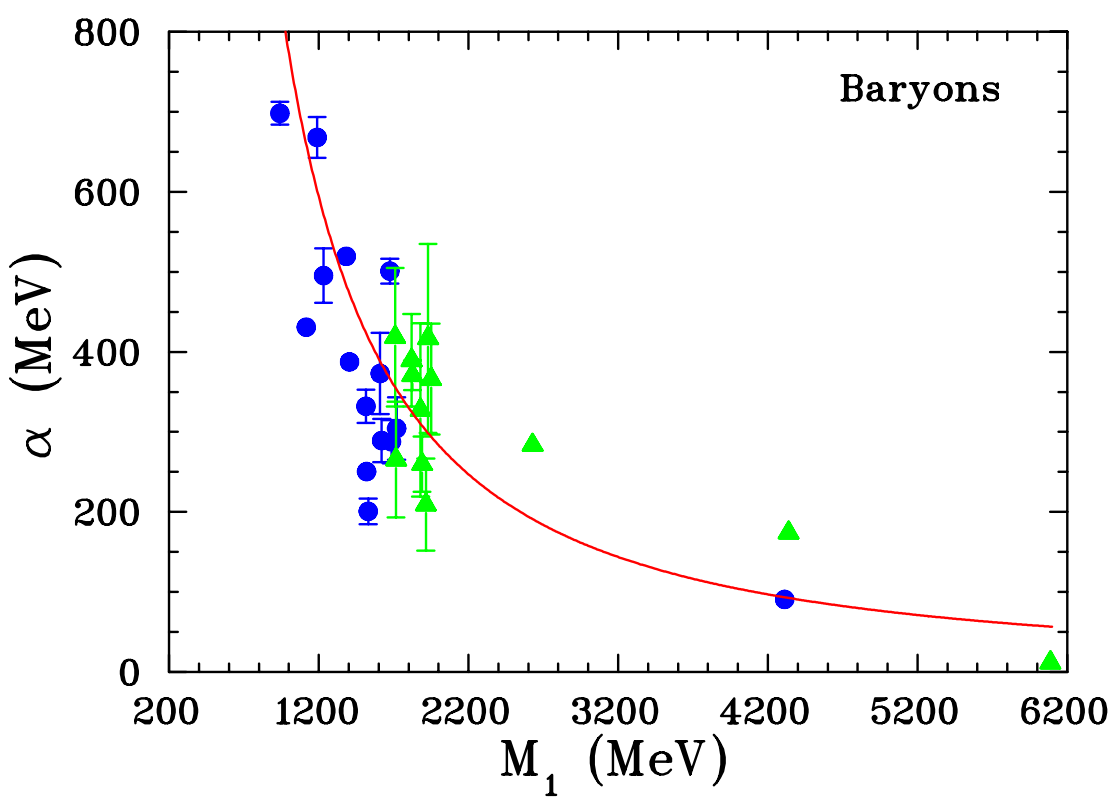}
    \caption{Baryon $\alpha$ vs $M_1$ power-equation fit [($\alpha=1.688\times10^7)M_1^{-1.446}~\mathrm{MeV}$]. Blue filled circles are from Part~I baryon sets~\cite{Roper:2024ovj}, green filled triangles are from this Part~II baryon sets~\cite{Roper:2025ldj}, and red curve is Parts~I and II fit to the power equation.
}
    \label{fig:BaMPEq}
\end{figure}

A plot of fit parameter $\alpha$ versus $M_1$ for all baryon sets in Part~I~\cite{Roper:2024ovj} and Part~II~\cite{Roper:2025ldj} roughly fits a power equation
(Fig.~\ref{fig:BaMPEq}). 

\section{UME Samples for Mesons}
\begin{figure}[htb!]
    \centering
    \includegraphics[width=0.4\linewidth]{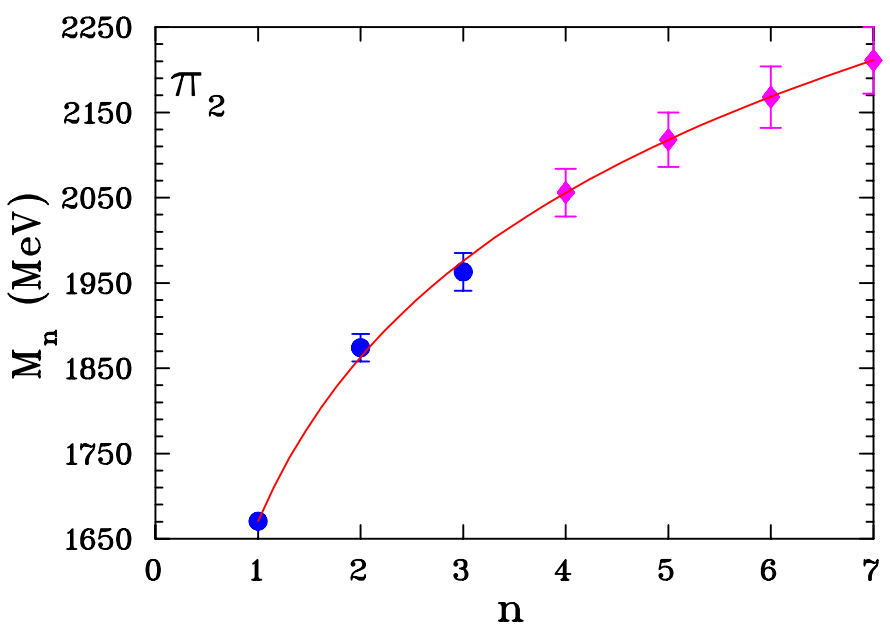}~~~
    \includegraphics[width=0.4\linewidth]{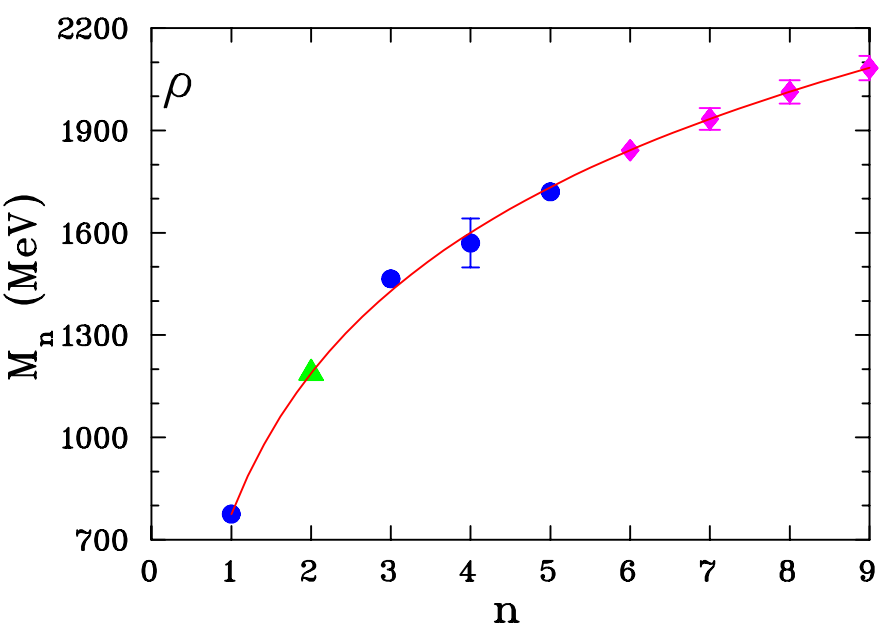}
    \caption{$\pi_2$ set (left) and $\rho$ set (right). 
    The notation is the same as in Fig.~\ref{fig:N1/2+}.
    Part~I~\cite{Roper:2024ovj}.
    }
    \label{fig:Pi2}
\end{figure}

\begin{figure}[htb!]
    \centering
    \includegraphics[width=0.4\linewidth]{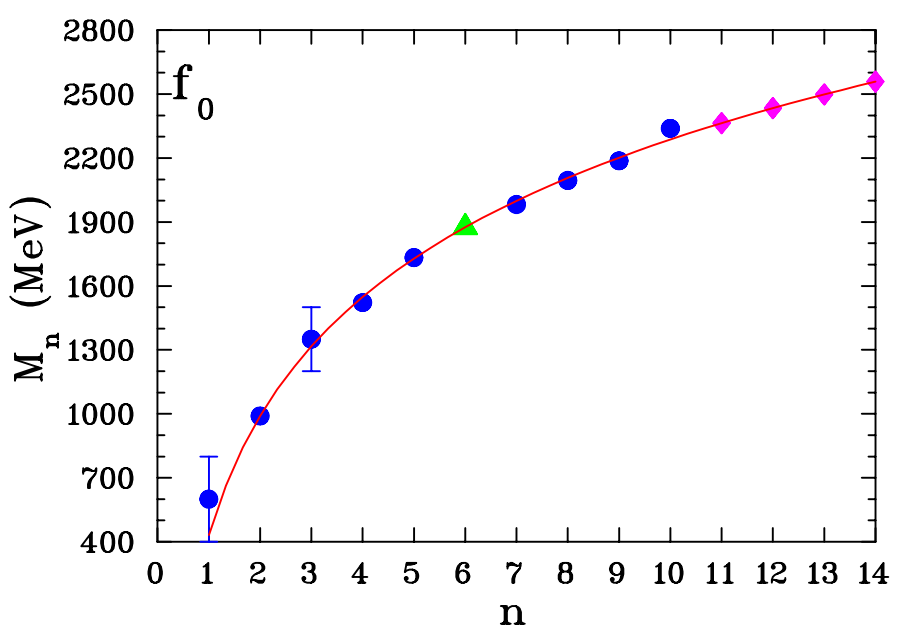}~~~
    \includegraphics[width=0.4\linewidth]{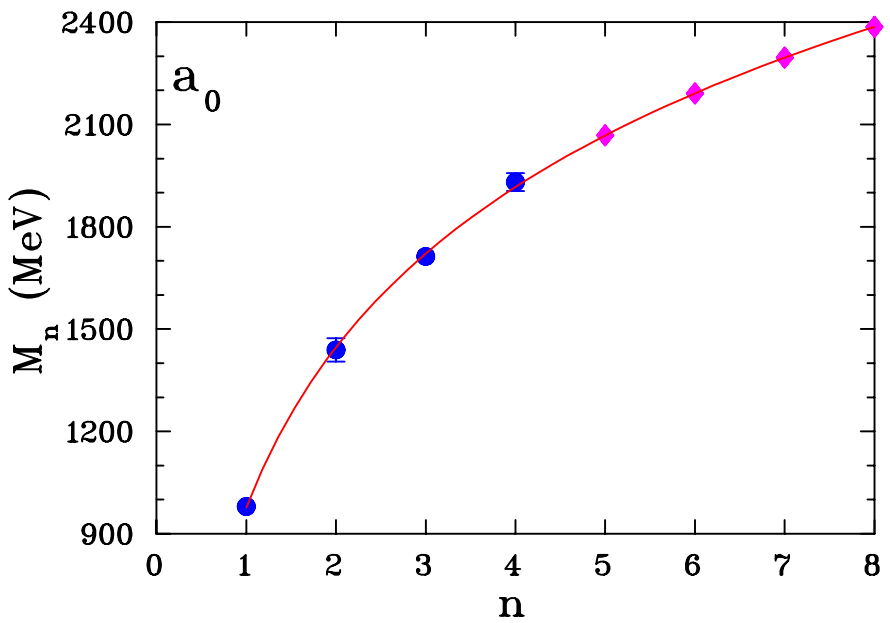}
    \caption{$f_0$ set (left) and $a_0$ set (right). 
    The notation is the same as in Fig.~\ref{fig:N1/2+}.
    Part~I~\cite{Roper:2024ovj}.
    }
    \label{fig:f0}
\end{figure}

\begin{figure}[htb!]
    \centering
    \includegraphics[width=0.4\linewidth]{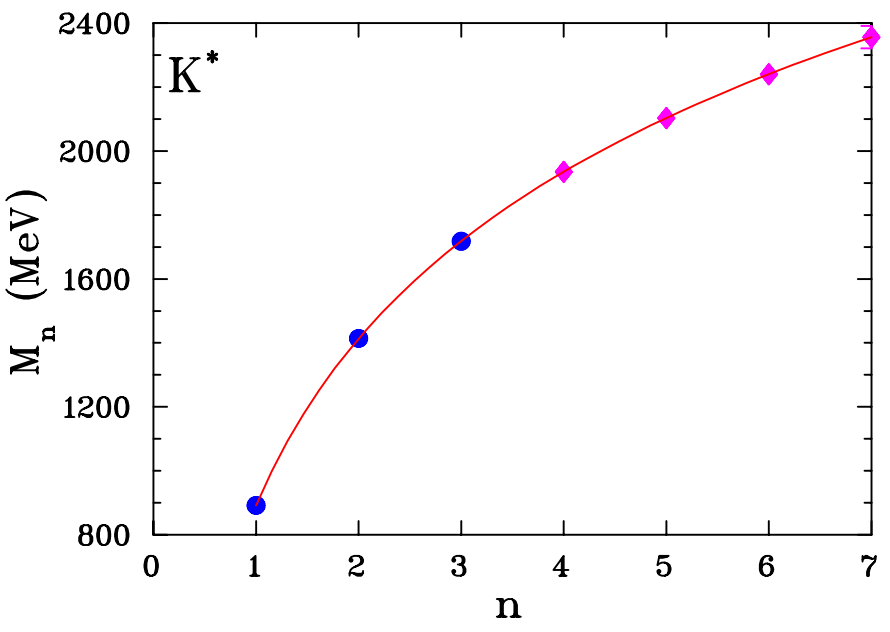}~~~
    \includegraphics[width=0.42\linewidth]{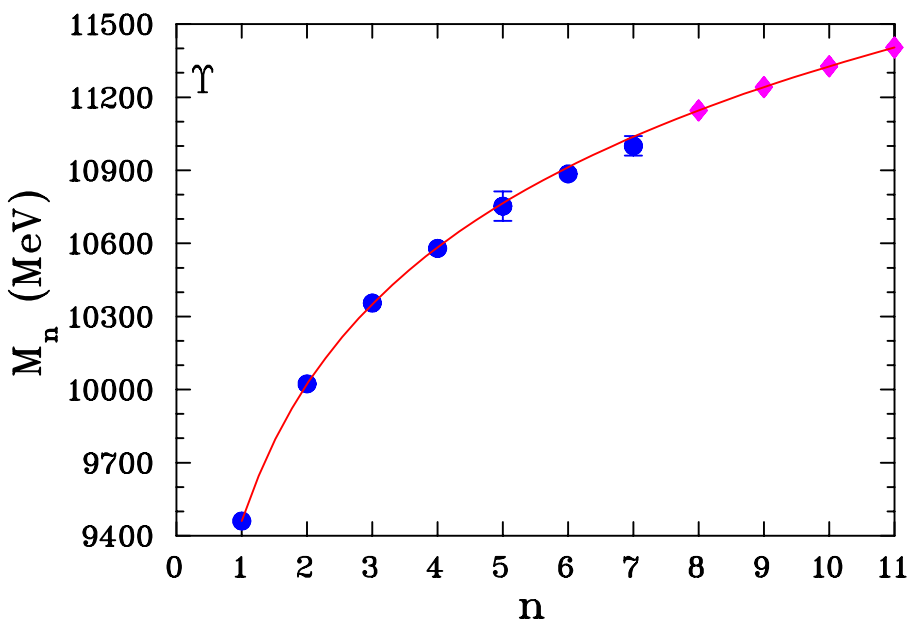}
    \caption{$K^*$ set (left) and $\Upsilon$ set (right). 
    The notation is the same as in Fig.~\ref{fig:N1/2+}.
    Part~I~\cite{Roper:2024ovj}.
    }
    \label{fig:K*}
\end{figure}

Samples for the meson case are given in Figs.~\ref{fig:Pi2} through \ref{fig:K*}.

\subsection{Systematics of Meson Excited-States Sets}
\begin{figure}[htb!]
    \centering
    \includegraphics[width=0.5\linewidth]{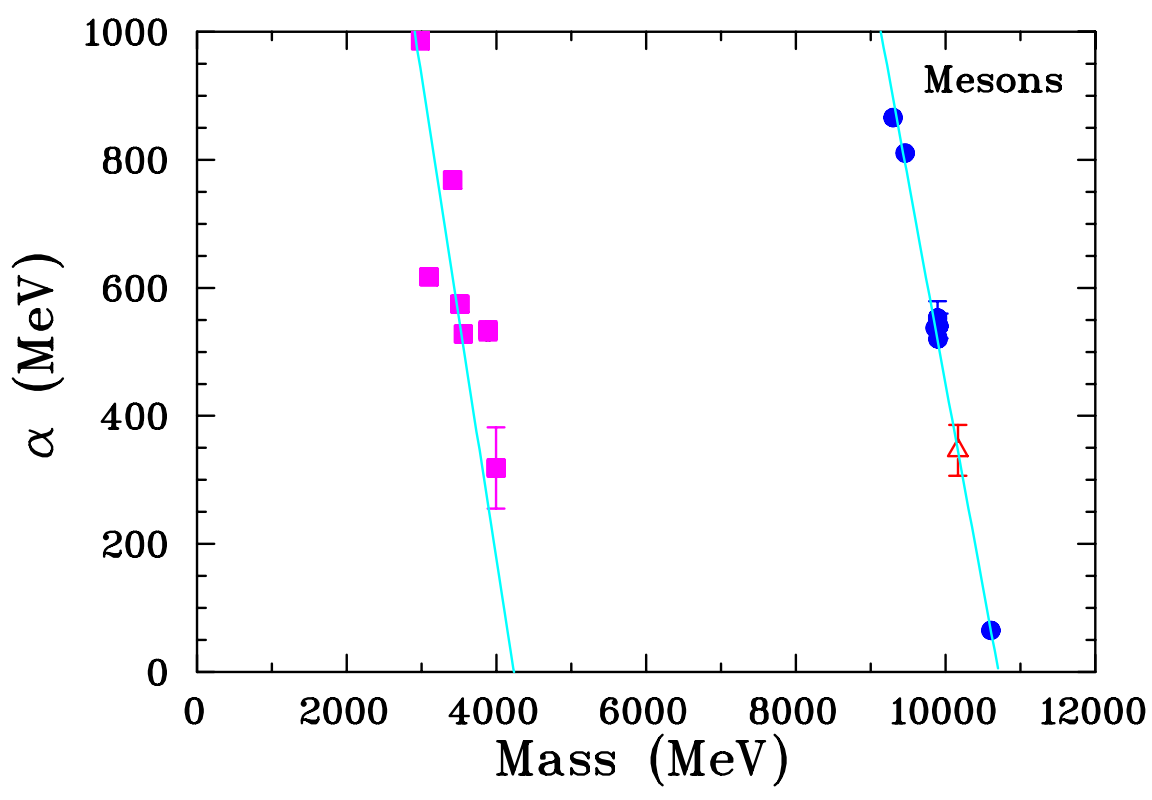}
    \caption{
    $c\bar{c}$: $\alpha=-(0.7539\pm0.0045)M_1+(3190.7\pm14.1)~\mathrm{MeV}$ (magenta filled squares).  
    $b\bar{b}$: $\alpha=-(0.6359\pm0.0028)M_1+(6809.3\pm27.4)~\mathrm{MeV}$ (blue filled circles).
    Line fits for $q\bar{q}$ and $s\bar{s}$ are not good. 
    The red open triangle shows $\Upsilon_2$ (
    Fig.~\ref{fig:Ups2}).
    Part~II~\cite{Roper:2025ldj}.
    }
    \label{fig:MaMLEw}
\end{figure}

The linear trend of $c\bar{c}$ and $b\bar{b}$ sets' $\alpha$ \textit{vs.} mass is shown in Fig.~\ref{fig:MaMLEw}. Because of its strong linearity, can be used to accurately estimate the parameter $\alpha$ for quantum states, for instance, for $b\bar{b}$ for which only the ground-state mass is known~\cite{ParticleDataGroup:2024cfk}, to be used to calculate the higher masses of the set. This illustrates an approximate inverse relationship of the fit parameter $\alpha$ with the ground-state mass. 
\begin{figure}[htb!]
    \centering
    \includegraphics[width=0.4\linewidth]{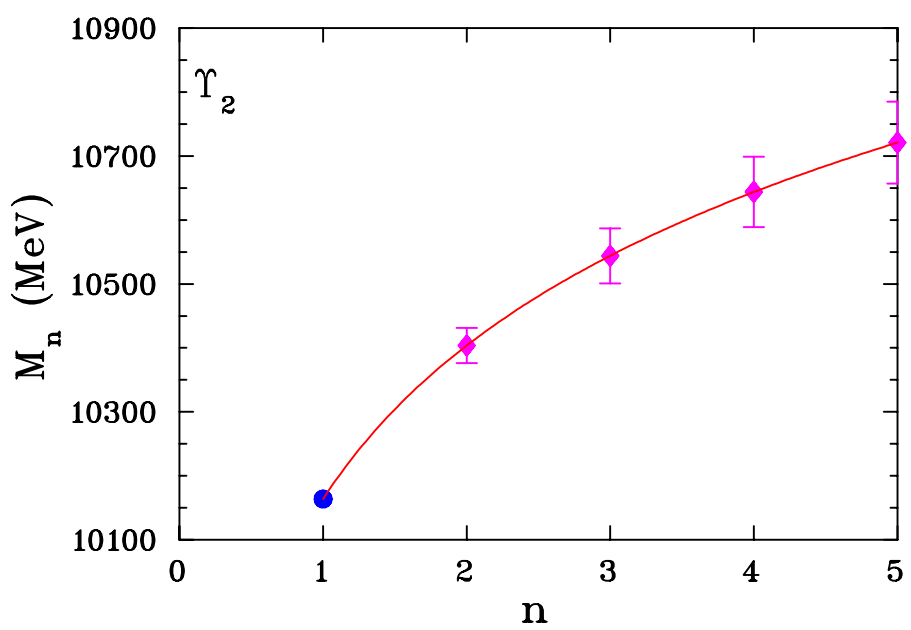}
    \caption{$\Upsilon_2$, a $b\bar{b}$ meson, has only one excited state 
    known~\cite{ParticleDataGroup:2024cfk}.
    The $b\bar{b}$-line fit (Fig.~\ref{fig:MaMLEw}) can be used to calculate its 
    $\alpha$. Part~II~\cite{Roper:2025ldj}. 
}
    \label{fig:Ups2}
\end{figure}

\section{Measurability of Predicted Excited States}
\begin{figure}[htb!]
    \centering
    \includegraphics[width=0.4\linewidth]{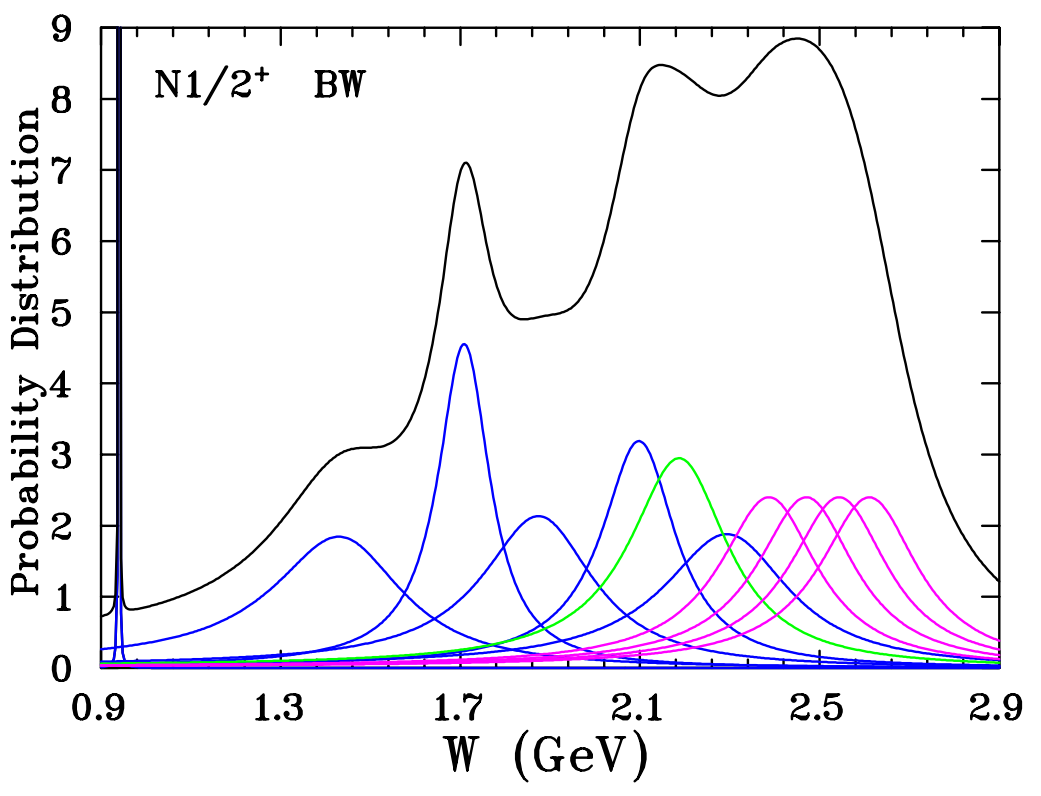}~~~
    \includegraphics[width=0.4\linewidth]{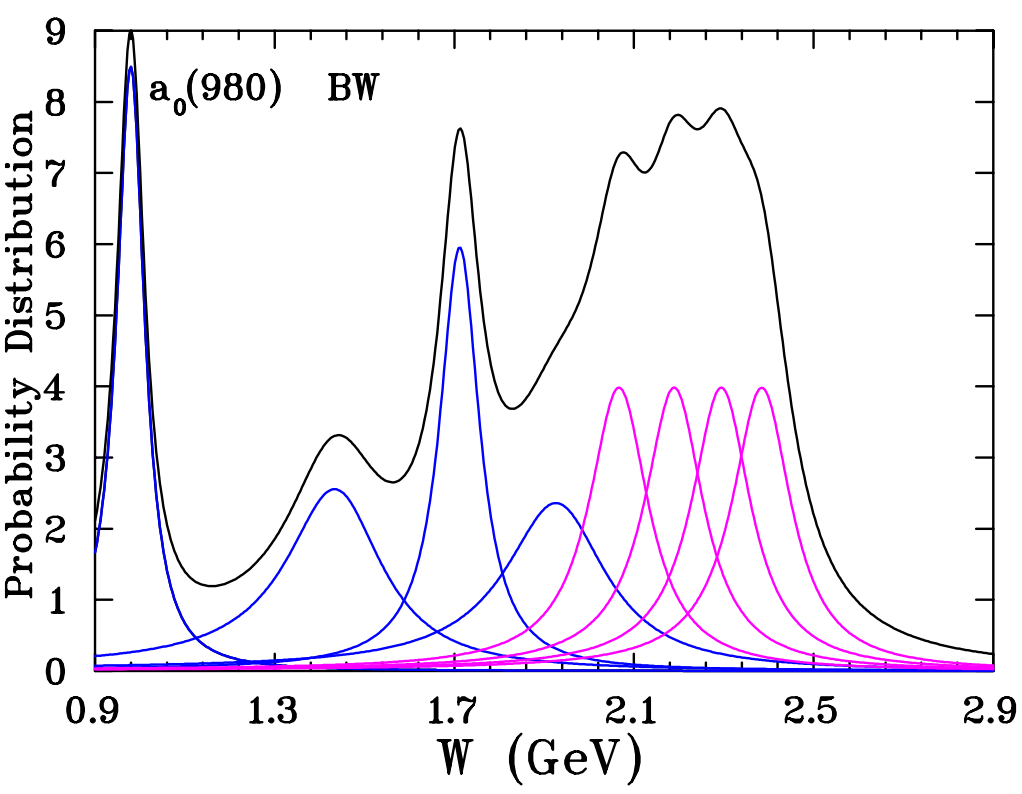}
    \includegraphics[width=0.4\linewidth]{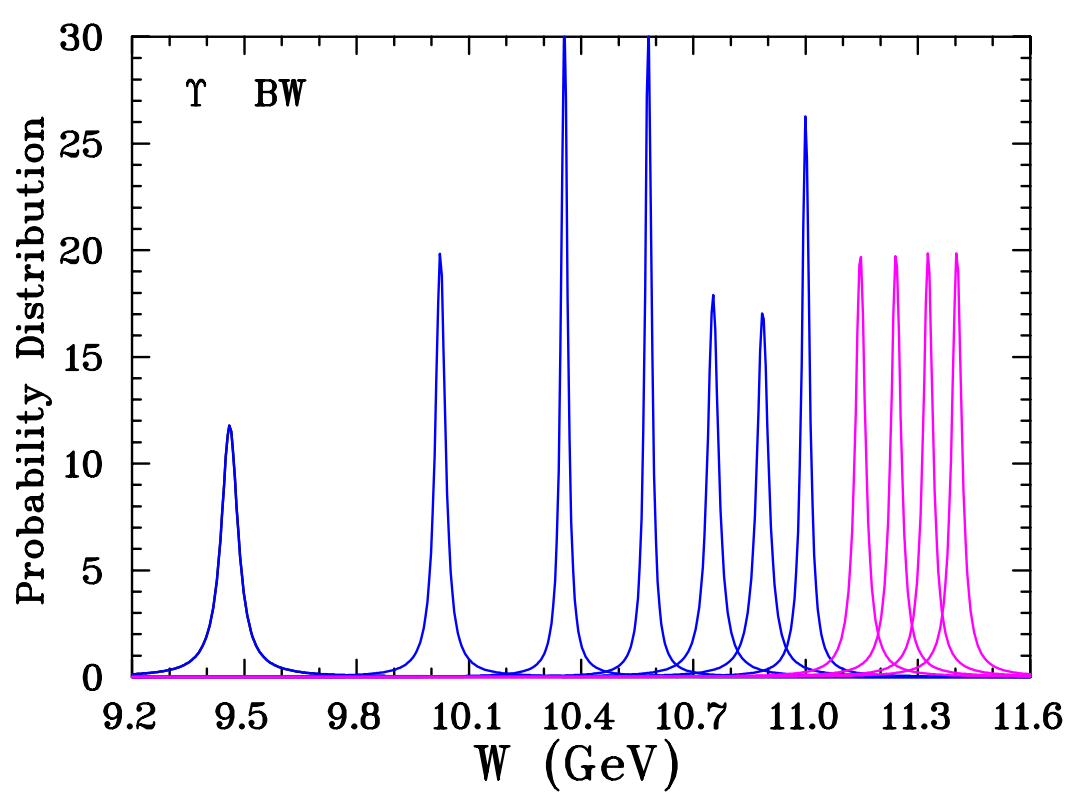}
    \caption{The final black curves are for the known, missed, and predicted states with average width of the preceding known states. $N1/2^+$ BW curves (top left) - possibly not measurable. $a_0$ (top right) BW curves - possibly measurable. $\Upsilon$ (bottom) - measurable. 
    The notation is the same as in Fig.~\ref{fig:N1/2+}. Part~II~\cite{Roper:2025ldj}.
    }
    \label{fig:N12pBW}
\end{figure}

\raggedright Use the accurate form of the BW resonance equation~\cite{Willenbrock:2025fgh}:
\begin{equation}
    P(s,M,\Gamma)=\frac{\Gamma}{2\pi}\frac{4M^2+\Gamma^2}{(s^2-M^2+\Gamma^2/4)^2+M^2\Gamma^2}
\end{equation}
to calculate the BW curves for predicted adjacent excited states to determine their overlap as an indication of their measurability. It illustrates that the ``bump hunting'' technology is problematic (Fig.~\ref{fig:N12pBW} (top)) to look for the predicted states 
because of the overlapping resonances.

\section{Potential Energy Curve}
\begin{figure}[htb!]
    \centering
    \includegraphics[width=0.4\linewidth]{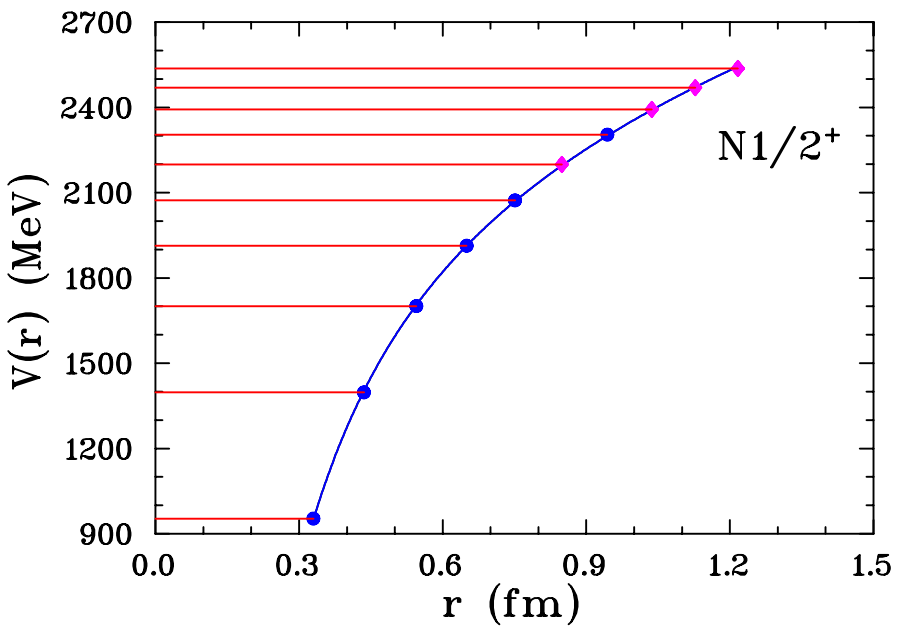}~~~
    \includegraphics[width=0.4\linewidth]{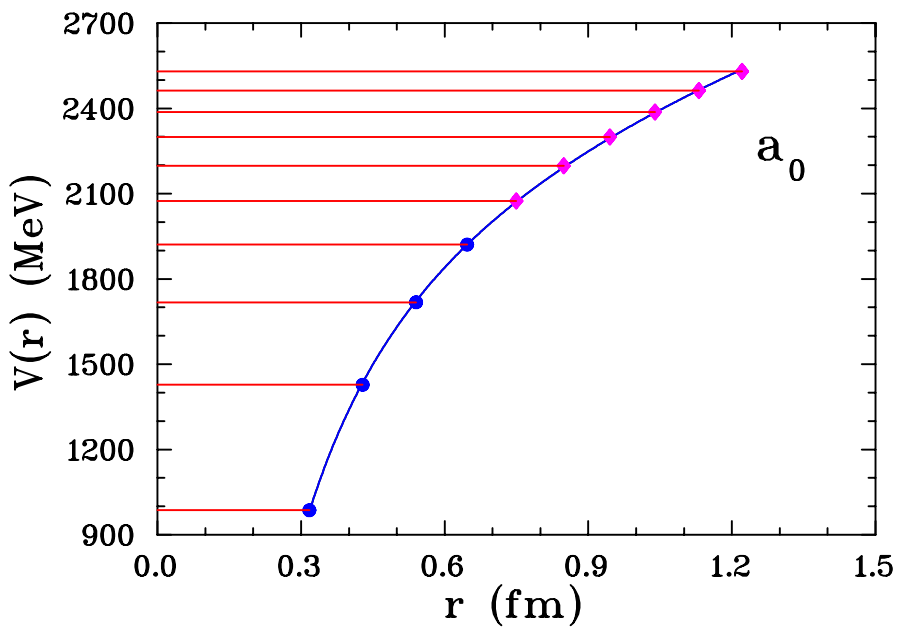}
    \caption{$N1/2^+$ baryon set (left) and $a_0$ meson set (right).
    The blue curves are the Cornell potential-energy function (Eq.~(\ref{eq:corn})) fits to the excited-states set's potential energy. Part~I~\cite{Roper:2024ovj}.
}
    \label{fig:N12pa0CP}
\end{figure}

The de~Broglie wavelength for a mass: $\lambda_n=\frac{h c}{M_n}=2\pi\frac{\hbar c}{M_n}=2\pi\frac{197.327}{M_n}~\mathrm{fm}$. Assume wavefunction radius, $r$, of excited state $n$: $r_n=n\frac{\lambda_n}{4}=\frac{\pi}{2}\frac{\hbar c}{M_n}$. A fit to the Cornell potential versus radius~\cite{Eichten:1974af}: 
\begin{equation}
   V(r)=-\frac{4}{3}A/r+Br+C
\label{eq:corn}
\end{equation}
is given in Fig.~\ref{fig:N12pa0CP}. Here, $r$ is the effective radius of the resonance state, $A$ is the QCD running coupling, $B$ is the QCD string tension, which controls intercepts and slopes of linear Regge trajectories, and $C$ is a constant. The first term of Eq.~(\ref{eq:corn}) represents a short distance, $r < 0.1~\mathrm{fm}$, and the second term corresponds to a long distance.

\section{Conclusion}
We have identified a UME that fits known equal-quantum excited-state sets and uses it to predict missing and higher-mass states. We discuss which of the predicted states are possibly measurable. We roughly define the wavefunction size and plot the excited states of a set versus the radius of the wavefunction and achieve a good fit of the wavefunction to the Cornell potential. Our UME should be of interest to experimentalists who are planning future accelerator experiments and to nuclear/particle theorists.

\acknowledgments
This work was supported in part by the U.~S.~Department of Energy, Office of Science, Office of Nuclear Physics under award No. DE--SC0016583.


\end{document}